\documentclass[sigplan,twocolumn,authorversion,screen]{acmart}
\renewcommand\footnotetextcopyrightpermission[1]{}
\usepackage{subcaption}
\usepackage[normalem]{ulem}
\usepackage{algorithm}
\usepackage{algpseudocode}
\usepackage{array}
\usepackage{booktabs}
\usepackage{caption}
\usepackage{colortbl}
\usepackage{enumerate}
\usepackage{enumitem}
\usepackage{fancyvrb}
\usepackage{graphicx}
\usepackage{listings}
\usepackage{makecell}
\usepackage{mathtools}
\usepackage{mdframed}
\usepackage{multirow}
\usepackage{pifont}
\usepackage{placeins}
\usepackage{rotating}
\usepackage{soul,color}
\usepackage{soul}
\usepackage{textcomp}
\usepackage{tikz}
\usepackage{ulem,lipsum}
\usepackage{url}
\usepackage{xcolor}
\usepackage{xspace}
\usepackage{subcaption}

\newcommand{\sgx}{SGX\xspace}
\newcommand{\sgxvo}{SGXv1\xspace}

\newcommand{\hashjoin}{HashJoin\xspace}
\newcommand{\btree}{B-Tree\xspace}
\newcommand{\bfs}{BFS\xspace}
\newcommand{\pagerank}{PageRank\xspace}
\newcommand{\svm}{SVM\xspace}

\newcommand{\openssl}{OpenSSL\xspace}

\newcommand{\kvs}{Key-Value\xspace}

\definecolor{LightCyan}{RGB}{211,211,211}

\newcommand{\one}{\ding{182}\xspace}
\newcommand{\two}{\ding{183}\xspace}
\newcommand{\three}{\ding{184}\xspace}
\newcommand{\four}{\ding{185}\xspace}
\newcommand{\five}{\ding{186}\xspace}
\newcommand{\six}{\ding{187}\xspace}
\newcommand{\seven}{\ding{188}\xspace}

\newcommand{\cmark}{\textcolor{blue}{\ding{51}}}%
\newcommand{\xmark}{\textcolor{red}{\ding{55}}}%

\newcommand{\ecall}{\texttt{ECALL}\xspace}
\newcommand{\ocall}{\texttt{OCALL}\xspace}

\newcommand{\sota}{state-of-the-art\xspace}

\lstdefinestyle{lstcode}{
language=C,
tabsize=4,
numbers=left,
showstringspaces=false,
basicstyle=\lstfont{black},
identifierstyle=\lstfont{black},
keywordstyle=\lstfont{blue},
numberstyle=\lstfont{black},
stringstyle=\lstfont{green},
commentstyle=\lstfont{brown!100},
emph={read,fsync,close,write,pread,pwrite,open,openat,lseek,CHUNK_DATA_SIZE_BYTES,HASH_SIZE,AES_KEY_SIZE
},
emphstyle={\lstfont{violet}},
breaklines=true,
frame=l,
captionpos=b,
    belowskip=1em,
aboveskip=1em
}

\mdfdefinestyle{mdframe_style}{%
    }

       \mdfdefinestyle{MyFrame}{%
       linecolor=black,
       outerlinewidth=2pt,
       skipabove=4pt,
       skipbelow=14pt,
       innertopmargin=4pt,
       innerbottommargin=4pt,
       innerrightmargin=4pt,
      innerleftmargin=4pt,
        leftmargin = 4pt,
        rightmargin = 4pt,
       backgroundcolor=gray!10!white,
       frametitlerule=true,
       nobreak=false,
       align=center
           }

\lstdefinestyle{cppstyle}{
    language=C++,
    basicstyle=\ttfamily\footnotesize,
    keywordstyle=\color{blue}\bfseries,
    commentstyle=\color{green!60!black},
    stringstyle=\color{orange},
    numbers=left,
    numberstyle=\tiny\color{gray},
    stepnumber=1,
    numbersep=0.2mm,
    backgroundcolor=\color{white},
    showspaces=false,
    showstringspaces=false,
    showtabs=false,
    frame=lines, 
    framerule=0.2pt, 
    rulecolor=\color{black},
    tabsize=4,
    captionpos=b,
    breaklines=true,
    breakatwhitespace=false,
    escapeinside={\%*}{*)},
    morekeywords={constexpr, decrypt},
    float
}

\lstdefinestyle{modcpp}{
    language=C++,
    basicstyle=\ttfamily\footnotesize,
    keywordstyle=\color{blue}\bfseries,
    commentstyle=\color{green!60!black},
    stringstyle=\color{orange},
    backgroundcolor=\color{white},
    showspaces=false,
    showstringspaces=false,
    showtabs=false,
    frame=lines, 
    framerule=0.2pt, 
    rulecolor=\color{black},
    tabsize=4,
    captionpos=b,
    breaklines=true,
    breakatwhitespace=false,
    escapeinside={\%*}{*)},
    morekeywords={constexpr, decrypt},
    float
}

\newcommand{\sepblock}{
\medskip
\noindent
}

\newcommand{\stopandcopy}{stop-and-copy\xspace}
\newcommand{\precopy}{pre-copy\xspace}
\newcommand{\postcopy}{post-copy\xspace}
\newcommand{\migsgx}{\textit{MigSGX}\xspace}

\newcommand{\tdxstar}{TDX$^*$}\xspace

\newcommand{\savethread}{\texttt{save\_th}\xspace}
\newcommand{\restorethread}{\texttt{restore\_th}\xspace}
\newcommand{\backupbuff}{\texttt{BBuff}\xspace}
\newcommand{\metabuff}{\texttt{MBuff}\xspace}

\newcommand{\criu}{CRIU\xspace}
\newcommand{\optmigmgr}{OptMgr\xspace}

\newcommand{\userfaultfd}{{userfaultfd}\xspace}
\newcommand{\masterkey}{$M_k$\xspace}

\newcommand{\savevec}{\texttt{save\_vec}\xspace}
\newcommand{\restorevec}{\texttt{restore\_vec}\xspace}
\newcommand{\mallocarray}{\texttt{MemArr}\xspace}
\newcommand{\restorechunk}{\texttt{\_\_process\_access()}\xspace}

\definecolor{Gray}{gray}{0.85}
\newcolumntype{C}[1]{>{\centering\arraybackslash}p{#1}}
\newcolumntype{G}[1]{>{\centering\arraybackslash\columncolor{Gray}}p{#1}}
\newcolumntype{L}[1]{>{\raggedright\arraybackslash}p{#1}}
\newcolumntype{R}[1]{>{\raggedleft\arraybackslash}p{#1}}
\newcolumntype{D}[1]{>{\centering\arraybackslash} m{#1}}

\newcommand{\trustedLib}{{\methodname}Lib\xspace}
\newcommand{\optmigvo}{{\methodname}-V1\xspace}
\newcommand{\optmigvt}{{\methodname}-V2\xspace}

\newcommand{\ncite}[1]{\citeauthor{#1}\cite{#1}}

\newcommand{\methodname}{ConstMig\xspace}

\setlist[enumerate]{itemsep=0pt, leftmargin=11pt, topsep=2pt}
\begin{document}

\title{\textit{\methodname}: Enabling Secure Live Migration of Large Intel SGX-based applications }

\begin{abstract}
Cloud service providers are adopting Trusted Execution Environments, or TEEs, to provide hardware-guaranteed security to applications running on remote, untrusted data centers. However, migrating such applications still uses the decade-old stop-and-copy-based method, which introduces large downtimes. Modern live-migration approaches such as \textit{pre-copy} and \textit{post-copy} do not work for such applications due to limitations imposed by the TEE.

We propose \textit{ConstMig}, a near-zero downtime live-migration mechanism for large memory footprint TEE-based applications. 
\methodname is fully compatible with containers, virtual machines (VMs) and microVMs. Our prototype, built on Intel SGX, a TEE solution from Intel, has a near-zero downtime irrespective of enclave size. ConstMig reduces the total downtime by 77-96\% for a suite of SGX applications with multi-GB memory footprints compared to state-of-the-art TEE-based migration, MigSGX.
\end{abstract}

\author{Sandeep Kumar} 
 \affiliation{ 
\institution{School of Information Technology\\IIT Delhi}
\city{New Delhi}
\country{India}
}
\email{sandeep.kumar@cse.iitd.ac.in}

\author{Abhisek Panda} 
 \affiliation{ 
\institution{Department of Computer Science and Engineering, IIT Delhi}
\city{New Delhi}
\country{India}
}
\email{abhisek.panda@cse.iitd.ac.in}

\author{Smruti R. Sarangi} 
    \affiliation{ 
    \institution{Department of Computer Science and Engineering, IIT Delhi}
\city{New Delhi}
\country{India}
}
\email{srsarangi@cse.iitd.ac.in}

\maketitle

\section{Introduction}
\label{sec:intro}
Cloud service providers, or CSPs, are adopting trusted execution environments, or TEEs, to
execute security-sensitive applications~\cite{cloudsec1, cloudsec2, cloudsec3,haven,stealthdb_sql_sgx_db,cryptsqlite_sgx_db,enclavedb_sgx_db,azue_always_encrypted_db,pribate_ml_tf_secexec,privacy_preserving_secexec,tfencrypted_secexec,salom_dnn_sgx_secexec,crypto_nn_design_secexec} on
remote, untrusted cloud nodes. In a TEE setting, the hardware ensures the application's security. The hypervisor and operating system (OS) are assumed to be untrusted (\S\ref{sec:bg_sgx}). Leading CSPs such as Microsoft and Google provide~\cite{google_cloud_offering,micrsoft_cloud_offering} several TEE options such as Intel SGX~\cite{sgx}, Intel TDX~\cite{intel_tdx}, AMD SEV~\cite{amd_sev}, ARM TrustZone~\cite{arm_trustzone} (with Confidential Compute Architecture or CCA planned~\cite{arm_cca}), and RISC-V Keystone~\cite{keystone}. 

\begin{figure}
        \centering
        \includegraphics[width=.9\linewidth]{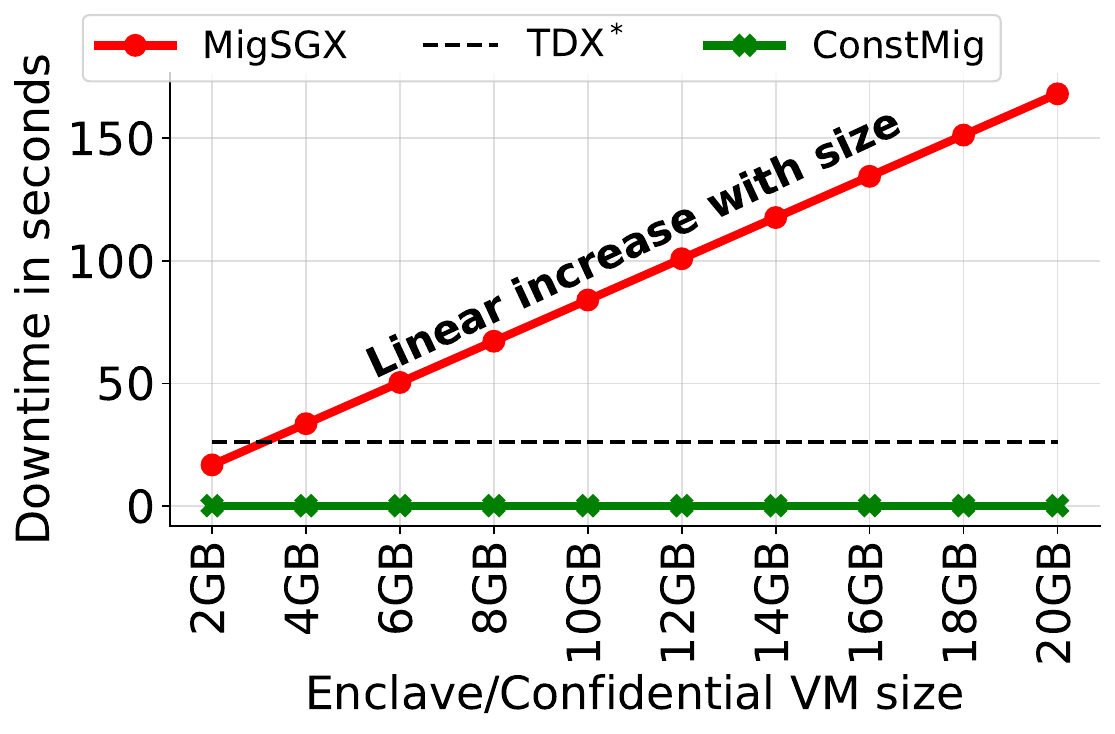}
        \caption{\methodname ensures a non-zero total downtime while migrating an \sgx-based application, irrespective of its size. TDX enables a hardware-based live migration but has long \textit{boot/respond} time. }
        \label{fig:teaser}
\end{figure}

Deploying TEEs in data centers presents unique challenges in migration of applications from one machine to another. CSPs rely on efficient \textit{live migration} to provide high availability, load balancing, efficient resource management, failure tolerance, seamless upgradation, and code patching~\cite{migimp1, migimp2, migimp3, migimp4}.
In a non-TEE setting, CSPs rely on modern live migration mechanisms such as \precopy~\cite{orig_pre_copy} and \postcopy~\cite{orig_post_copy} employed by hypervisor or a migration manager   (\S\ref{sec:motiv_scalibity}) to achieve a near-zero downtime during migration~\cite{orig_pre_copy,orig_post_copy}. However, these techniques are not available in a TEE setting as they require access to either \textit{userspace fault handling} (\userfaultfd) or \textit{tracking page writes} (using \texttt{DIRTY} bits in the page table) which are prohibited by the TEE hardware (see Section~\ref{sec:motiv_scalibity})~\cite{amd_sev_precopy_postcopy_not_supported, arm_trustzone_userfaultfd,arm_trustzone_userfaultfd_systex22} as the operating system (OS), hypervisor, or any other migration manager is not a part of the trusted computing base (TCB).

Recognizing this challenge, TEE vendors integrated hardware support to enable live migrations of TEE applications. A \textit{migration manager} typically orchestrates the migration and uses the API exposed by the hardware to handle the secure data pages to TEE applications -- which otherwise would be inaccessible (or ineligible) to the migration manager due to TEE security guarantees. AMD enables migration of secure VMs using Migration Agents -- special VMs that provide migration keys, enforce guest policies, and form part of the VM’s TCB~\cite{amd_migration}. Intel TDX uses a small VM known as MigTD that assists the untrusted VMM in accessing the secure state (after encryption). ARM TrustZone does not support live migration; however, the proposed ARM CCA will have hardware-enabled live migration~\cite{arm_cca}.
Intel SGX, which allows for TEE enablement at the per-application level, has no hardware support for migration.  
Hence, prior work in enabling migration for SGX relied on the classical
\stopandcopy method for migrating SGX
enclaves~\cite{migsgx_mm,mig_containers_mm,mig_persistent_mm,mig_teender_hsm_mm,mig_replicatee_consensus_mm}.
In the \stopandcopy method, an executing application is
paused, a checkpoint is taken, and the checkpoint is transferred to the
destination machine, where it is subsequently resumed.
This approach has a large downtime, which is proportional to the 
memory footprint (see Figure~\ref{fig:teaser}). \textit{Given that Intel has planned support for SGX till Intel Xeon 7 -- Diamond Rapids\cite{dmr_sgx}, we believe that missing live migration support is a key omission which this paper aims to solve.}
Secure VMs provided by Intel TDX, which do have hardware support for live migration, suffer from malicious co-hosted applications, a large TCB, and a high memory encryption cost even in
the unsecure portions of an application (performance overheads). SGX has a small TCB, low initialization time, and no overheads for the unsecure portion (\S\ref{sec:sgx_vs_tdx}).


We present \textit{\methodname}, a novel live migration mechanism for \sgx enclaves with a near-zero downtime that is independent of the memory footprint of the application. 
\methodname requires no additional hardware and has far lower performance overheads than the state-of-the-art (see Figure~\ref{fig:teaser}). 
%
\methodname addresses a fundamental challenge in enabling live migration for enclave-based systems: a fine-grained tracking of memory accesses on the destination machine to ensure that the restored application reads valid data from the source machine -- instead of an invalid, all-zero page (\S\ref{sec:fine_grained_tracking}). 
We propose two access tracking mechanisms: \one based on fine-grained page fault tracking ( \S\ref{sec:fault_tracker}) using kernel modifications and \two based on fine-grained pointer access tracking (\S\ref{sec:access_tracker}).
In addition to providing all the security guarantees of \sgx, \methodname also provides
a freshness guarantee during the migration process, which ensures that an attacker cannot replay old encrypted enclave data during migration (\S\ref{sec:security_analysis}).

Our key contributions are as follows:
\begin{enumerate}
    \item We identify the key challenges faced while adopting modern migration schemes for enclaves.
    \item We propose a novel migration scheme that addresses these challenges and enables near-zero downtime while migrating large-sized \sgx applications. To our knowledge, {\it we are the first to implement this
and show its benefits}.
    \item We propose novel optimizations during the migration process that reduce the total migration time and reduce the cost of migration.
    \item We evaluate our solution using a set of widely used \sgx benchmarks. Our scheme outperforms the current \sota~\cite{migsgx_mm} by 77-96\% for a suite of Intel SGX applications that have multi-GB memory footprints.
\end{enumerate}
\section{Background \& Motivation}
\label{sec:background_and_motivation}

Here, we discuss some relevant background for the paper.

\subsection{Intel Software Guard eXtensions or SGX }
\label{sec:bg_sgx}
Intel Software Guard Extensions (\sgx) enable secure execution of applications on untrusted machines by creating a secure sandbox called an \textit{enclave}~\cite{intelsgxexplained}. These enclaves are isolated from even privileged entities like the OS, hypervisor, and system administrators. At boot, \sgx reserves a portion of main memory
which is hardware-managed and encrypted using a hardware key inaccessible to any software component. The usable part of this reserved space is called the \textit{Enclave Page Cache} or \textit{EPC} (the rest is used by \sgx to store metadata). 
The latest version of \sgx, referred to as scalable \sgx, supports an EPC of up to 512\,GB~\cite{epc_1tb, epc_64GB}, compared to the earlier version (\sgxvo) with a usable EPC of only 92\,MB~\cite{securelease,securefs,intelsgxsdk}. 

\noindent
\textit{Memory management:} \sgx ensures an enclave's integrity by validating the signature of a loaded enclave~\cite{intelsgxexplained} at initialization time. However, this process can become costly for enclaves with large memory footprints, as all pages must be provisioned upfront. 
For instance, as per our experiments, initializing a 20\,GB enclave takes  $\approx 30$ seconds. To address this, Intel introduced \textit{Enclave Dynamic   Memory Management} (EDMM)~\cite{sgx-hasp}, which allows developers to specify a minimal set of \textit{committed pages} during initialization, while additional \textit{uncommitted pages} can be added dynamically~\cite{tug_of_war}. In principle, EDMM enables an enclave requiring 512\,GB to start with a single committed 4\,KB page, though this flexibility may introduce performance overhead depending on the workload~\cite{tug_of_war, edmm_managemetn_systor}.

\noindent
	extit{\sgx execution framework:} \sgx provides security at the application level. Here, an application is partitioned into \textit{secure} and \textit{unsecure} components. The secure component executes within an enclave, incurring performance overheads due to security guarantees, whereas the unsecure component runs conventionally without overhead (see Figure~\ref{fig:scoping}). Interaction between the two occurs via \textit{enclave calls} (\ecall) from the unsecure side and \textit{outside calls} (\ocall) from the enclave~\cite{intelsgxexplained}.

\subsection{Intel Trusted Domain Extension or TDX}
\label{sec:bg_intel_tdx}


Intel Trusted Domain eXtension or TDX is the next-generation TEE offering from Intel which introduces the concept of a hardware-protected confidential virtual machine (CVM)~\cite{confidentiality_vms_explained,gramine_tdx}-- also known as Trusted Domain (or TD)-- instead of just a single application. TDX allows unmodified applications to execute inside a hardware-protected VM (although the OS and the hypervisor must be modified to handle TDX specific interrupts~\cite{intel_tdx_demystified,confidentiality_vms_explained}). TDX enables live migration of a CVM with hardware support.

Intel TDX supports dynamic memory management through the \textit{Dynamic Physical Address Metadata Table} (PAMT), which stores metadata essential for securely managing physical memory assigned to Trust Domains (TDs). While the entire DRAM is encrypted using Intel Multi-Key Total Memory Encryption (MKTME), TDX enables dynamic allocation and deallocation of PAMT entries, allowing memory pages to be added to or removed from a TD at runtime, enabling full DRAM encryption and integrity protection without huge initialization overheads.

\subsection{Intel SGX vs Intel TDX}
\label{sec:sgx_vs_tdx}
Intel TDX is the latest TEE offering from Intel which aims to secure the whole VM and has hardware support for a live migration. TDX is the natural choice for deploying an application with TEE support with minimal development effort. However, SGX brings its own advantages that are either lacking or missing in TDX.

\subsubsection{Execution Overhead}

As discussed, Intel \sgx divides an application into secure and unsecure portions. The secure portion executes in an enclave inside \sgx and incurs the overhead due to security guarantees of \sgx. The untrusted portion, however, suffers no additional performance overhead. 
However, Intel TDX does not follow such a partitioned approach. An application executes completely within a CVM and all of its data is protected by the TDX module~\cite{confidentiality_vms_explained}.

As per \ncite{sgx_vs_tdx_vs_sev}, PyTorch inference executed on SGX with the LibOS Gramine~\cite{gramine} demonstrates improvements in performance of up to 78.6\% compared to Intel TDX,
and 32.25\% compared to AMD SEV-SNP, respectively. Furthermore, \ncite{sgxgauge} illustrates that programs executing natively on SGX, after the partition of secure and insecure code, outperform the performance of those using LibOS Gramine by up to 3$\%$.

\sepblock
{\it CVM switch overhead:} In a TEE solution, frequent transitions between secure and unsecure worlds and
memory encryption are the main causes of performance overheads. \ncite{confidentiality_vms_explained} show that workloads that require frequent VM exits can suffer performance slowdowns of up to 240\% and 472\%  on AMD SEV and Intel TDX, respectively, compared to traditional VMs. The additional overhead in TDX is mainly due to the involvement of the TDX module in handling VM exits, which adds extra processing
steps~\cite{confidentiality_vms_explained}.

\begin{figure}[t]
    \centering
    \includegraphics[width=.9\linewidth]{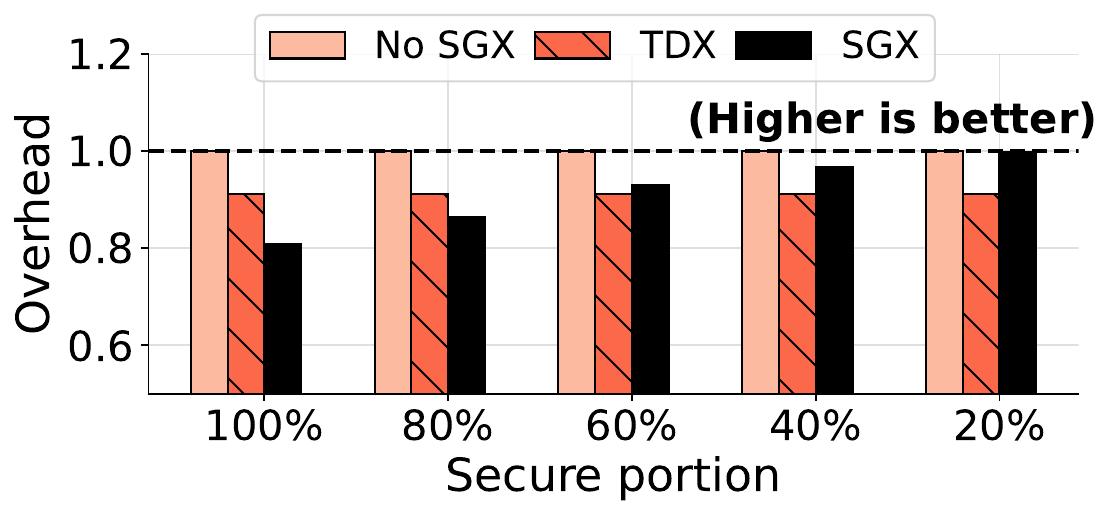}
    \caption{Performance of \kvs with No-SGX, TDX, and \sgx for varying amount of secure portion of the allocated data. }
    \label{fig:sgx_flexibile_perf}
\end{figure}

\sepblock
	extbf{SGX Advantage:}
\sgx's flexible execution framework can address the performance
concerns due to security measures. With \sgx, the performance overhead in terms
of memory encryption and costly enclave entry and exit are relevant only for the secure
portion of the application (\S\ref{sec:bg_sgx}). The untrusted part
executes in a conventional manner with no performance loss.
Figure~\ref{fig:sgx_flexibile_perf} shows the end-to-end performance of \kvs workload for varying amount of sensitive data -- for 100\% everything executes in \sgx and, say, for 0\% only 60\% data is in \sgx, rest in the untrusted region outside \sgx.
With 100\% of data as secure, \sgx has a performance overhead of $\approx$12\% compared to TDX. However, with 20\% of data as secure, \sgx outperforms TDX by $\approx$ 9\% and approaches the performance of No-SGX.

\subsubsection{TCB Size:} The TCB (Trusted Computing Base) of TDX includes the TDX-enabled processor with
Virtualization Technology (VT), Multi-key Total Memory Encryption (MKTME) and
SGX~\cite{intel_tdx_demystified}, along with TDX Module ($\approx$20\,K LOC~\cite{tdx_intel_online}), and the entire guest
OS and software stacks within the
CVM~\cite{intel_tdx_demystified}. Such a large TCB provides a wide attack surface and makes Intel TDX inherently susceptible to errors and vulnerabilities\cite{confidentiality_vms_explained}. \ncite{confidentiality_vms_explained} report that 54\% of the identified vulnerabilities related to Intel TDX are associated with the underlying firmware. 
SGX's TCB consists of the processor, the microcode and firmware that provide security, and the secure part of an application's code. The untrusted portion of the application and the host operating system are \textit{not} part of the TCB. 


\subsubsection{Initialization overhead}
\begin{figure}[t]
    \centering
    \includegraphics[width=.8\linewidth]{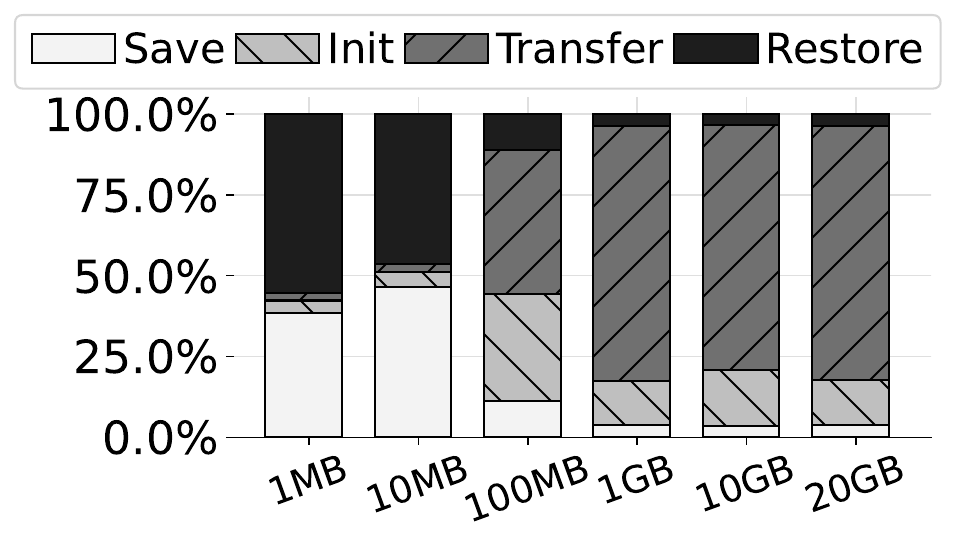}
    \caption{For \stopandcopy, for large enclaves, the total migration time is dominated by the enclave initialization time on the destination machine and the total transfer time.}
    \label{fig:motiv_snapshot_time}
\end{figure}

After the network transfer time, the time to initialize an enclave/CVM affects the migration time the most ($17\%$, see Figure~\ref{fig:motiv_snapshot_time}).
With \sgx, the initialization time of an enclave is proportional to the number of committed
pages -- a low count of committed pages reduces the initialization time, but may incur performance overhead~\cite{tug_of_war,edmm_managemetn_systor}. 
TDX uses PAMT to dynamically add 4\,KB pages after booting. We observed that a TDX CVM on an Emerald Rapid Machine (\S\ref{sec:eval_setup}) takes $\approx 26$ seconds to boot up, irrespective of the allocated memory (similar observation made by~\ncite{confidentiality_vms_explained}).

\subsection{Migration methods and Challenges with SGX}
\label{sec:motiv_scalibity}
The performance of a migration method is evaluated using two metrics: \textit{total downtime} -- the period during which the application is inactive -- and \textit{total migration} time -- the duration from migration start to completion~\cite{metrics_mig_criu,migsgx_mm}.

\subsubsection{Stop-and-Copy Migration} 
\label{sec:bg_stopncopy}
In the \stopandcopy migration approach~\cite{orig_pre_copy}, the application is paused on the source machine while its state (e.g., registers, memory, open files, network) is transferred to the destination machine. It resumes execution only after the transfer completes. As a result, both total migration time and downtime scale are proportional to the application's memory footprint.

The state-of-the-art migration mechanism for \sgx~\cite{migsgx_mm} is based on
\stopandcopy, which has a total downtime proportional to the size of the enclave
(\S\ref{sec:bg_stopncopy}). In our experiments, we observed that the total time
to migrate an enclave of size (mostly heap size) 1\,GB is $\approx$ 8 seconds,
which increases to $\approx$160 seconds when the enclave size is 20 GB, an
increase of 20$\times$. This represents a significant scaling issue, as the size of the
enclave with the latest version of \sgx can be scaled up to 512\,GB.
 
Figure~\ref{fig:motiv_snapshot_time} shows the four phases of \stopandcopy migration: save, transfer, initialize and restore. For small enclave sizes ($\approx$ 10 MB), the total migration time is
dominated by the save and restore phases. However, for larger enclave sizes ($>$100 MB), the network transfer phase dominates the total migration time (up to 96\% for 10 GB+ enclaves).




\begin{figure}[t]
    \centering
    \subfloat[Pre-copy migration \label{fig:pre_copy}]{
    \includegraphics[width=.7\linewidth]{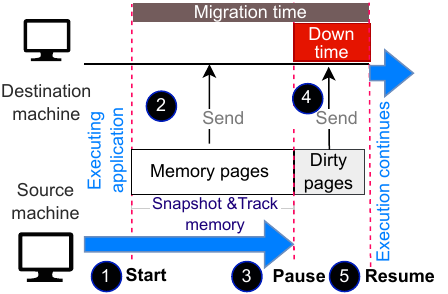}
    }
    
    \subfloat[Post-copy migration \label{fig:post_copy}]{
        \includegraphics[width=.6\linewidth]{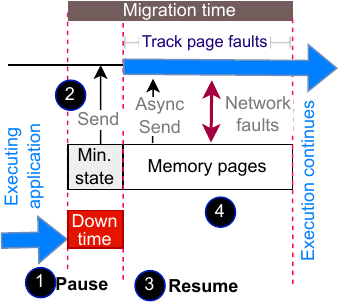}
    } 
    \caption{\precopy and \postcopy migration mechanisms. The total downtime is independent of the memory footprint. \textbf{Start},\textbf{Pause}, and \textbf{Resume} indicate the start of migration, pausing, and resuming of the application, respectively.}
    \label{fig:migration_schemes}
\end{figure}

\subsubsection{Pre-copy Migration}
\label{sec:bg_precopy}
As shown in Figure~\ref{fig:pre_copy}, \precopy migration proceeds as follows:
\one Snapshot the application's state and \textit{track} subsequent write operations.
\two Transfer the snapshot to the destination. \three Identify and track modified (dirty) pages. If their count is below a threshold, \four pause the application and transfer the dirty pages else send the dirty pages again and go back to step three. \five Resume the application on the destination. 
Here, total migration time scales with memory footprint, while downtime depends on the number of dirty pages in the final transfer.

\noindent
\textbf{Challenge in adopting for SGX:} Tracking an application's address space to identify dirty pages is essential for the \textit{pre-copy} migration method. Non-TEE migration approaches typically rely on the page table entry’s \textit{dirty} bit, which is automatically set by the hardware upon a write to a data page~\cite{criu_track_mem,criu_track_mem_lwn,soft_dirty_pte}. Prior work~\cite{wang_leaky,van_dirty_bit_without_pagefault} has shown that dirty bits are also set for SGX-backed pages, enabling the tracking of write operations within enclaves. However, these findings were based on experiments conducted on the Intel Skylake i7-6700, a desktop-class processor released almost a decade ago~\cite{skylake_6700}. Since then, Intel has discontinued SGX support on desktop CPUs, restricting it to server-class processors~\cite{sgx_only_on_xeon}, and has introduced several security enhancements to mitigate SGX-related attacks~\cite{sgx_fix_1,sgx_fix_2,aex_notify}.

In our experiments using the latest Xeon-based servers (Icelake and Emerald-Rapids), we observed that \textit{\ul{dirty bits are not set for SGX-backed pages}}, even after repeated write operations (see Appendix~\ref{appendix:dirty_bit} for a micro-benchmark to validate this claim).

\subsubsection{Post-Copy Migration}
\label{sec:bg_postcopy}
In post-copy migration (Figure~\ref{fig:post_copy}), \one the application is first paused on the source machine, and a minimal state (CPU registers, non-pageable memory~\cite{orig_post_copy,orig_pre_copy}) is sent to the destination machine. \two The application is then resumed on the destination, while the remaining state is transferred in the background. \three During the transfer, memory accesses to pages still on the source machine trigger \textit{network page faults}, which are transparently handled by the OS and migration mechanism. Here, total migration time scales with memory footprint, but downtime remains constant. However, performance may degrade during migration due to costly network faults.

\noindent
\textbf{Challenge in adopting for SGX:} 
Tracking memory accesses of a restored application is critical for the \textit{post-copy} migration. In non-TEE environments, migration frameworks use the \userfaultfd mechanism~\cite{userfaultfd} to intercept page faults. With {\userfaultfd}, the OS forwards page faults to the migration manager, which then fetches the missing page from the source machine and services the fault. However, {\userfaultfd} supports only anonymous memory, huge pages, and shared memory regions~\cite{userfaultfd}. As \sgx enclave memory is device-backed via \texttt{/dev/isgx}~\cite{userfaultfd_lwn,userfaultfd_lwn_nextstep,userfaultfd,vma_can_userfaultfd}, \userfaultfd cannot track memory accesses within enclaves.

{

\begin{figure*}[t]
  \centering
  \subfloat[Scope of \methodname. \label{fig:scoping}]{
  \includegraphics[width=.2\linewidth]{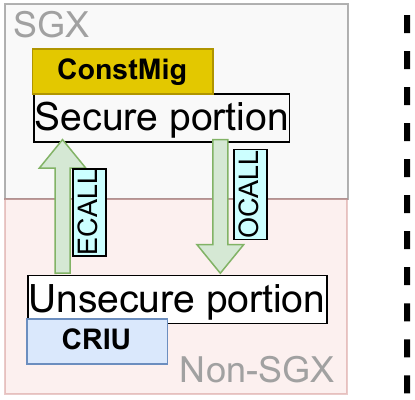}
  }
  \subfloat[A high-level overview of the working of \methodname. \label{fig:optmig_high_level}]{
  \includegraphics[width=.7\linewidth]{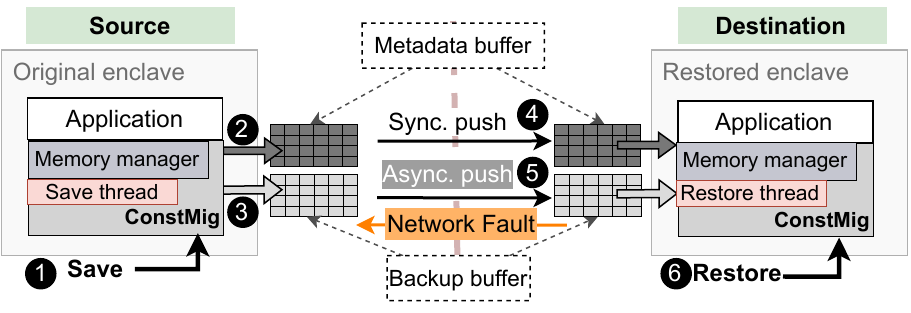}
  }
  \caption{\methodname handles the migration of the secure portion of an application. The rest is handled by \criu.}
\end{figure*}

\section{Design}
\label{sec:design}


In this section, we discuss the design of \methodname.
As shown in Figure~\ref{fig:scoping}, \methodname handles the migration of the secure portion of the application. \criu handles the migration of the untrusted portion.
The primary design goals of \methodname are as follows: 
\begin{enumerate}
  \item \textbf{Low downtime:} Enable live migration for large \sgx enclaves with minimal total downtime
  \item \textbf{Performance overhead:} Minimize performance degradation due to migration.
  \item \textbf{Security:} Maintain confidentiality, freshness, and integrity of the data at all times during migration.
\end{enumerate}

\noindent
We maintain the following invariants during the migration operation:

\noindent \textbf{Invariant 1:} It is never possible that a page fault with virtual address $V$ on the destination machine is served with the data of a page with virtual address $V'$ ($V \ne V'$). \label{invar:only_heap}
    
\noindent \textbf{Invariant 2:} Once the migration key is transferred, the source \emph{cannot} be resumed. \label{invar:mater_key_linearizability}

\noindent \textbf{Invariant 3:} A page is ``saved'' and ``restored'', at most once. \label{invar:use_of_bit_vectors}

\noindent \textbf{Invariant 4:} Migration components executing outside \sgx never see the plain text data. \label{invar:confidentiality}

\noindent \textbf{Invariant 5:} The migration key (\masterkey) is a function of the \texttt{PUF} of the source machine, boot time, current time, and the output of a \texttt{TRNG}. {A replay attack is thus not possible (\S~\ref{sec:security_analysis}).}
\label{invar:migration_key}

\subsection{Scope}
As discussed earlier, modern TEEs support hardware-assisted live migration, with the notable exception of Intel SGX. Given Intel’s plans to continue SGX support on future server-class processors, enabling live migration for large SGX enclaves remains an important and timely challenge. Our focus is primarily on heap memory, which constitutes the majority of an application's memory footprint~\cite{invisipage}. In contrast, other segments such as \textit{data} and \textit{BSS} typically range from a few KBs to MBs and are \textit{always} transferred synchronously as part of the minimal state during migration.

Furthermore, we do not discuss migration of \textit{sealed data}. \sgx provides a way to seal data using a hardware-bound key. \ncite{mig_persistent_mm} proposed a method to migrate sealed data by encrypting all the sealed data with a software-generated key on the source machine and resealing it on the destination machine after transfer and decryption. This solution can be easily incorporated into \methodname.

\subsection{Threat Model}
\label{sec:threat_model}
We adopt the same security assumptions as Intel SGX as prior work in this domain. The operating system and other software components outside the enclave are \textit{not} trusted. They do not have access to enclave plaintext data or the cryptographic keys used during migration.

In SGX, the OS manages the page tables of secure applications. A malicious OS can manipulate the \texttt{PRESENT} bit in page table entries (PTEs) to infer the application's page access patterns~\cite{controlledchannelattacks}. Such side-channel attacks fall outside the threat model of SGX and are beyond the scope of this work (see \S\ref{sec:security_analysis}). Migration is aborted immediately upon detection of any security violation. 


\subsection{High-Level Overview}
\label{sec:design_high_level_ops}
There are two main components in \methodname: the \textit{migration manager} (CRIU)
that orchestrates the entire migration process and a trusted \sgx library
(\textit{\trustedLib}), which is linked to the secure application (see Figure~\ref{fig:scoping}). Its routines
have access to the secure memory space of the application running in the
enclave.

\sepblock
\textbf{Start migration:} As shown in Figure~\ref{fig:optmig_high_level}, 
at the beginning of the migration, \one the migration manager pauses the application and
sends the \texttt{save} signal to \trustedLib. 
The migration manager then resumes a single thread for \trustedLib to do its work ($save\,\,thread$).
\two \trustedLib generates a master key \masterkey to be used during migration. It
then saves the data segment, BSS segment and \optmigmgr metadata (information about the allocated regions) in an untrusted metadata buffer (\metabuff). 
\three A thread starts to save heap data in a separate untrusted backup buffer (\backupbuff). All these
operations are done after performing encryptions and adding integrity checks (\S\ref{sec:design_async_save} and \S\ref{sec:design_async_restore}).

\sepblock
\textbf{Handshake:} \four The migration manager sends a signal to spawn a fresh enclave on the destination machine after the start of the migration operation.
Once the enclave is initialized, \methodname first performs a remote attestation~\cite{remote_attestation} to verify that only a valid enclave with  \trustedLib is running on the machine, thereby establishing the ``chain of trust'' on the destination machine.
Once the enclave is validated, \methodname transfers \masterkey to the destination machine using a secure key transfer protocol that ensures that the key is only transferred to the intended destination machine~\cite{remote_attestation}.
\textit{\uline{The original enclave cannot be resumed after this (see Invariant 2).}}
If migration fails for any reason, the application instance is lost on both machines.
\five Following this, the destination machine receives the encrypted \metabuff.

\sepblock
\textbf{Start Restore:}
\six \trustedLib on the destination machine, decrypts \metabuff using \masterkey and reads the encrypted {\tt BSS} segment, data segment, and memory manager's metadata. 
\six The migration manager on the destination machine
sends the \texttt{restore} signal to \trustedLib. \seven A background thread starts reading data from \backupbuff and starts restoring the enclave's heap ($restore\,\,thread$). Since \backupbuff is stored in the untrusted region, it is protected by \userfaultfd~\cite{userfaultfd}, and the restore thread is guaranteed to read valid pages. 

\sepblock
\textbf{Cleanup:} Once the restoration is done, \trustedLib zeros all the allocated pages on the source machine and kills the application -- preventing {\it forking} attacks (\S\ref{sec:security_analysis}). At any given time, \textit{an application runs on only one system}.

\subsection{Implementation Details}
\label{sec:implementation_details}

\subsubsection{Migration Manager}
\label{sec:migration_manager_criu}
We use the widely popular \textit{Checkpoint/Restore In Userspace } or \criu~\cite{criu,criu_in_mem_docker,metrics_mig_criu,migsgx_mm} package as our migration manager. \criu supports both the \precopy and \postcopy live migration methods. 
In \methodname, \criu is \textit{not} a part of the trusted code base, and never sees the migrating enclave data in plaintext. 

To support SGX enclave migration, we extend CRIU with enclave-aware functionality. A \texttt{save} signal initiates a state save process and marks the beginning of a migration, while a \texttt{restore} signal triggers restoration when resumed on the destination machine. The signals are captured and processed by \trustedLib.

\sepblock
	extbf{Encrypted buffers}: We utilize two encrypted buffers during migration: a metadata buffer (\metabuff) that stores the BSS, data segments, and \methodname-specific metadata, and a backup buffer (\backupbuff) to store the encrypted heap contents. Both buffers are allocated in untrusted memory outside the SGX enclave.


Due to its relatively small size, \metabuff is transferred synchronously to
the destination at the beginning of migration.
In contrast, \textbf{\backupbuff} -- containing large encrypted heap data -- is transferred asynchronously. 
We integrate the \textit{userfaultfd} mechanism via CRIU to support on-demand page retrieval for \backupbuff. Since it resides in untrusted memory, read operations on the destination machine either access already-transferred data or trigger synchronous page fetches over the network~\cite{orig_post_copy}.


\sepblock
\textbf{A consistency checker:}
\criu handles the restoration of \backupbuff on the destination machine. However, \criu is not \sgx migration aware. To ensure correctness, we enforce two key constraints:
\begin{enumerate}
    \item \backupbuff is transferred only after the enclave state has been successfully saved on it.
    \item The restored enclave only reads restored data.
\end{enumerate}

To enforce these constraints, our consistency checker maintains three bit-vectors—internal \savevec, \savevec, and \restorevec of size equal to the number of 4\,KB pages in \backupbuff.
 The internal \savevec, allocated within the trusted region, tracks when a page is ready for transfer. A copy of this vector, \savevec, is maintained in the untrusted region for CRIU’s use. \restorevec, stored inside the enclave, indicates whether a page has been restored into enclave memory. All entries in \savevec and \restorevec are initialized to zero.
 


\noindent
	extit{Fine-grained Access tracking:} 
To ensure that a restored enclave accesses only valid pages, we enforce that a page read is permitted only if the corresponding bit in \restorevec is set. Otherwise, the access is paused until the page is fetched from the source machine. Such support is missing from \sgx as of now. We introduce two novel mechanisms to capture enclave memory accesses: access tracker and fault tracker. The access tracker automatically inserts memory access checks into the application's source code (using an LLVM pass), enabling fine-grained control. In contrast, the fault tracker leverages page table entries to detect access attempts at runtime.

Note that using page table entries to track leaks the memory access pattern. However, as discussed earlier (\S\ref{sec:threat_model}), such leakage is unavoidable in SGX, as the OS --an untrusted entity --manages the page tables (\S\ref{sec:threat_model}). We provide a detailed comparison of these approaches in Section~\ref{sec:fine_grained_tracking}.


\subsection{\methodname library (\trustedLib)}
\label{sec:trusted_library}

\begin{figure*}[t]
    \centering
    \begin{subfigure}[b]{.48\linewidth}
    \includegraphics[width=0.9\linewidth]{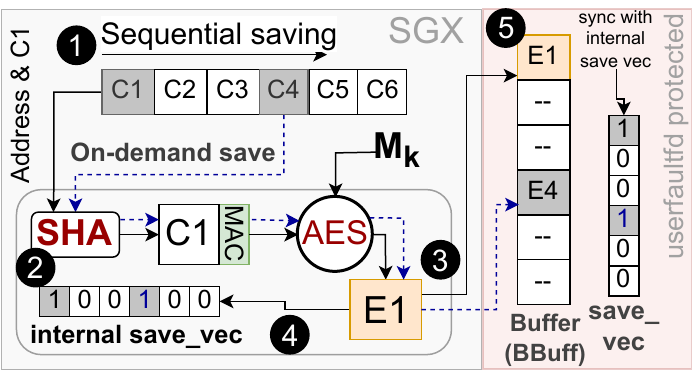}
    \caption{Asynchronous data saving.  }
    \label{fig:design_ecall_save}
    \end{subfigure}
    \begin{subfigure}[b]{.48\linewidth}
    \includegraphics[width=0.8\linewidth]{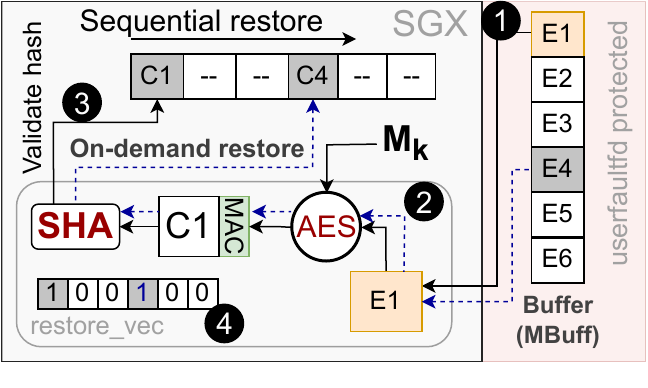}
    \caption{Asynchronous data restore.  }
    \label{fig:design_ecall_restore}
    \end{subfigure}
    \caption{Asynchronous data restoring. Note that \restorevec is allocated inside an enclave as there is no need on the destination machine for \criu to read it.}
\end{figure*}

To perform migration, \methodname requires knowledge of the application's allocated memory regions, along with mechanisms to save and restore them. We achieve this by statically linking \trustedLib to the SGX application, enabling support for memory region discovery and migration. \trustedLib executes inside \sgx and has access to all of the application's address space in plain text form. Its  main components are:

\subsubsection{Memory Manager} 
We implement a custom memory manager (\optmigmgr) by overriding the default \texttt{malloc()} and \texttt{free()} functions to intercept all allocation and deallocation requests. For each allocation, we record the address, size, and a unique ID in a structured array (\mallocarray). Upon deallocation, the corresponding entry is removed from \mallocarray. 
During migration, \mallocarray is encrypted and stored in \metabuff, which is transferred synchronously along with the enclave’s data and BSS segments to the destination machine.

\subsubsection{Save thread} 
\label{sec:design_async_save}
A dedicated save thread (\savethread) handles the \texttt{save} command initiated by CRIU and begins capturing the heap contents into \texttt{\backupbuff}. As shown in Figure~\ref{fig:design_ecall_save}, operating at a granularity of 4\,KB pages, {\savethread} performs the following steps for each page $P$ at virtual address $VA_p$: it concatenates the page contents $PC(P)$ with the address to form $PC(P) || VA_p$, and computes a hash $H$ over this concatenation. The page contents and hash are then encrypted using \texttt{\masterkey}, and the resulting encrypted packet is stored in \texttt{\backupbuff}. Once a page is saved, the corresponding bits in both \texttt{\savevec} and \texttt{internal} \texttt{\savevec} are updated to indicate readiness for transfer to the destination machine.

 
\subsubsection{Restore thread} 
\label{sec:design_async_restore}
A dedicated restore thread ({\restorethread}) handles the \texttt{restore} command initiated by CRIU and begins restoring heap contents from {\backupbuff}. Operating at a granularity of 4\,KB pages, the thread retrieves encrypted data from {\backupbuff}, decrypts it within SGX using {\masterkey}, and extracts the page contents ($PC(P)$) and hash ($H$). It then computes the expected hash ($H'$) and verifies that $H == H'$. If validation succeeds, the page is loaded into enclave memory; otherwise, a security violation is reported and migration is aborted (see Figure~\ref{fig:design_ecall_restore}). After successful restoration, {\restorethread} sets the corresponding bit in {\restorevec}, marking the page as restored. If the bit was already set, indicating a duplicate restoration attempt, a security violation is reported and migration is aborted (see Invariant~\ref{invar:use_of_bit_vectors}).

\noindent
\textit{Network Faults:} 
During restore, if the requested encrypted page is not available in {\backupbuff} (yet to be copied from the source machine), CRIU issues a network page fault to the source machine, and {\restorethread} enters a blocked state. Upon receiving the request, the source machine’s CRIU signals {\savethread} to save the page into {\backupbuff}. {\savethread} first checks the internal {\savevec} to verify whether the page has already been saved. If so, a security violation is reported. Otherwise, the page is encrypted and stored in {\backupbuff} (\S\ref{sec:design_async_save}), and CRIU transfers the encrypted page to the destination. Upon receipt, \restorethread validates the encrypted contents before restoring them into enclave memory.

\begin{figure}[h!]
\centering
\begin{minipage}{0.48\textwidth}
\begin{algorithm}[H]
\footnotesize
\caption{Prepare Encrypt Packet}
\label{algo:prepare_enc_packet}
\begin{algorithmic}[1]
\Procedure{PrepareEncryptPacket}{$P, VA_P, M_k$}
\State Check(\text{internal }\savevec)
\State $H \gets \text{Hash}(PC(P) \| VA_P)$ \Comment{Link page content with address}
\State $C \gets \text{Encrypt}(PC(P)\|H, M_k)$
\State Update (\savevec, internal \savevec)
\State \textbf{return} $C$
\EndProcedure
\end{algorithmic}
\end{algorithm}
\end{minipage}
\hfill
\begin{minipage}{0.48\textwidth}
\begin{algorithm}[H]
\footnotesize
\caption{Validate Encrypt Packet}
\label{algo:prepare_dec_packet}
\begin{algorithmic}[1]
\Procedure{ValidateEncryptPacket}{$C, VA, M_k$}
\State Check(\restorevec)
\State $PC(P)\|H \gets \text{Decrypt}(C, M_k)$
\State $H' \gets \text{Hash}(PC(P) \| VA)$
\If{$H = H'$}
    \State Update(\restorevec)
    \State \textbf{return} \textbf{true}
\Else
    \State \textbf{return} \textbf{false}
\EndIf
\EndProcedure
\end{algorithmic}
\end{algorithm}
\end{minipage}
\end{figure}


    


\section{Fine-Grained Access Tracking}
\label{sec:fine_grained_tracking}
A key challenge in enclave migration is that both the restored enclave and SGX hardware are \textit{not} migration-aware. Upon restoration, the enclave can access any addressable memory region; however, if it touches a page not yet restored by \restorethread, it triggers a page fault. Instead of retrieving the original encrypted page from \backupbuff, the fault handler allocates a fresh zeroed page -- leading to a \textit{use-before-restore} issue. As discussed in Section~\ref{sec:motiv_scalibity}, this cannot be mitigated using \userfaultfd, since SGX pages are device-backed (via $/dev/isgx$) and unsupported by userfaultfd in the Linux kernel~\cite{userfaultfd_lwn_nextstep,userfaultfd_lwn,userfaultfd_modes_kernel,vma_can_userfaultfd}.

To address this issue, we propose two novel schemes: the \textit{fault tracker} and the \textit{\methodname access tracker}. They ensure that a restored enclave reads valid data instead of a fresh all-zero page.

\subsection{Fault Tracker}
\label{sec:fault_tracker}

\begin{figure}[t]
    \centering
  \includegraphics[width=.6\linewidth]{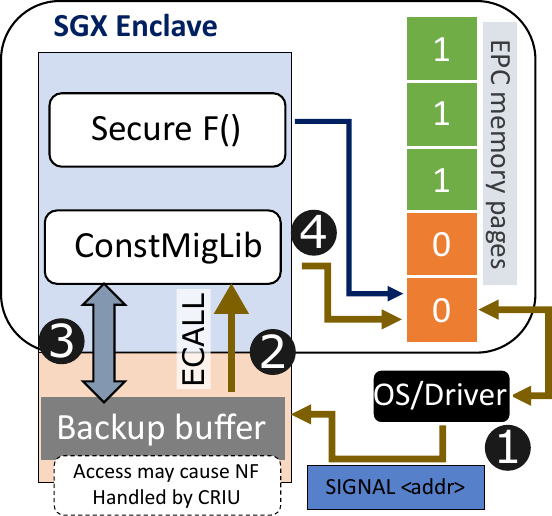}
  \caption{Working of \textit{Fault Tracker} to ensure a consistent memory access in the restored enclave}
  \label{fig:fault_tracker}
\end{figure}

The first approach (see Figure~\ref{fig:fault_tracker}) modifies the SGX page fault handler to return restored data instead of zeroed pages on the destination machine. Since enclave page tables are managed by the OS, a fault is triggered when the enclave accesses an unallocated page. By default, the SGX fault handler returns a zeroed page; we extend it to fetch the corresponding encrypted page from \backupbuff (which may cause a network fault that is transparently handled by \criu).
However, the page must be decrypted before the application can use it. As the fault handler operates in kernel space and cannot decrypt enclave data, it invokes enclave services to perform decryption, validation, and placement of the restored page.


\sepblock
\textbf{Fault capture and restoration:} 
As illustrated in Figure~\ref{fig:fault_tracker}, we extend the SGX driver to signal the application upon a page fault. Upon receiving the signal, the application performs an \ecall into the \trustedLib enclave with the faulting address. \trustedLib first checks whether the fault is due to an ongoing restoration by \restorethread. If not, it retrieves the encrypted page from \backupbuff, decrypts and verifies its integrity, and writes the restored data to the faulting address. This ensures that the faulting thread receives the correct data instead of a zeroed page. Importantly, each enclave page triggers exactly one signal -- either during restoration or due to a fault -- maintaining Invariant 3.


\sepblock
{\bf Implementation: } 
To intercept page accesses in the restored enclave, we leverage protection bits in page table entries~\cite{prot_none}, setting \texttt{PROT\_NONE} on committed pages via \texttt{mprotect}~\cite{mprotect}. We modify the \texttt{sig\_handler} in the trusted SGX SDK (PSW~\cite{intelsgxsdk}) to handle protection faults and signal the application after clearing the protection bit. A dedicated trusted thread, \texttt{constmig\_fault\_thread}, is introduced in the SGX SDK to handle these signals. This thread performs an \ecall into \trustedLib to decrypt, verify, and restore the faulting page. A separate thread is necessary since the faulting thread remains blocked until the fault is resolved; reusing it would crash the enclave.

\subsection{Access Tracker}
\label{sec:access_tracker}
\begin{figure}[t]
    \centering
  \includegraphics[width=.6\linewidth]{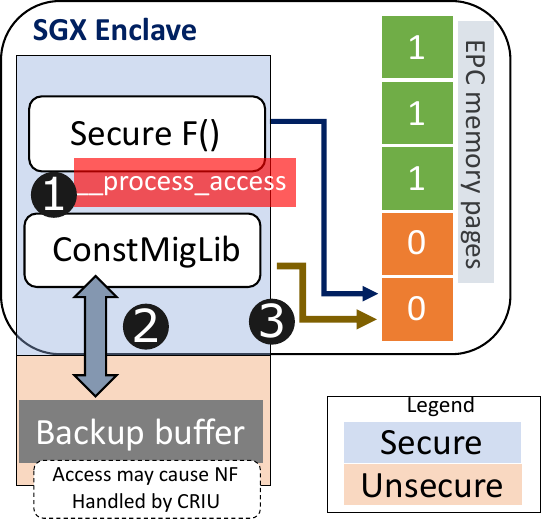}
  \caption{Working of \textit{Access Tracker} to ensure a consistent memory access in the restored enclave}
  \label{fig:access_tracker}
\end{figure}

The second approach introduces a runtime check before each heap pointer access in the restored enclave (see Figure~\ref{fig:access_tracker}). During an active migration, this check consults \restorevec to verify whether the target pages have been restored from \backupbuff. Listing~\ref{node_traversal} shows an example using the \restorechunk function, which executes entirely within the enclave and does not require an \ecall.
Once migration is complete (all the pages are restored -- indicated by a global flag), the memory check is replaced with \texttt{return 0;}.

The function \restorechunk takes a pointer address and access size, performs a linear scan of \mallocarray -- the array that contains information about all the allocated regions -- to locate the corresponding memory region, and identifies the required pages in \backupbuff. While linear scans suffice for a small number of regions, a radix tree can optimize lookups at scale. It then checks \restorevec to determine if all required pages are already restored. If not, it reads the encrypted pages from \backupbuff (may cause network faults, handled transparently by \criu), decrypts, and restores them.

\begin{lstlisting}[keywords={__process_access},caption={List traversal showing the added heap pointer access checks},label={node_traversal},escapechar=|]
void traverseList(){
  struct Node* pos = head;
  __process_access(pos, sizeof(struct Node));  
  while(pos!=NULL){ |\label{line:while_traversal}|
    __process_access(pos, sizeof(struct Node));  
    // use pos->val
    pos = pos->next;
  }
}
  \end{lstlisting}

\sepblock
	extbf{Implementation:} We implement a Clang-based source-to-source transformer~\cite{clang_pass_1,clang_pass_2} to insert memory access checks before heap pointer dereferences. This technique, widely used for pointer safety~\cite{softbound,cheri,pointercheck,heapcheck}, ensures that accessed pages have been restored. Our transformer extends Clang’s \texttt{RecursiveASTVisitor} to traverse the abstract syntax tree (AST)~\cite{ast}, which represents source code constructs such as variables and functions. For instance, the heap variable \texttt{pos} in Listing~\ref{node_traversal} appears as a \texttt{DeclRefExpr} node, containing its name and type. Using this information, the transformer inserts a call to \restorechunk before each heap pointer access.

\subsection{Fault Tracker v.s. Access Tracker}
\label{sec:eval_fault_tracker_vs_access_tracker}

The choice between the two trackers depends on the extent of changes a cloud vendor is able to implement in its system. 

\sepblock
{\bf Access to Source code}
Fault Tracker requires no enclave source code modification (apart from adding a
signal handler in the untrusted part of the application).
On the other hand, Access Tracker requires enclave source code modification to add checks before pointer accesses.

\sepblock
{\bf Page restoration}
Fault tracker restoration is expensive because it involves sending a signal and making an \ecall for each page restoration, whereas Access tracker is less overhead-prone because it does not require any signals or {\ecall}s (as the checks are done inside the enclave). 

Figure~\ref{fig:fault_tracker_vs_access_tracker} shows the time taken by each step in Fault and Access Trackers (resp.) for  4\,KB data and 4\,MB data, respectively.
For 4\,KB and 4\,MB of data, Access Tracker outperforms Fault Tracker by $35\%$ and $93\%$, respectively. The performance improvement for Access Tracker grows with an increase in the size of the data to be restored. Fault Tracker operates at a 4\,KB granularity; hence, restoring 4\,MB of data will require $\approx$ 1,024 faults, signals, and {\ecall}s. Whereas Access Tracker can potentially use the \mallocarray to restore the complete 4\,MB region using a single \ecall. The cost to restore a page (read from \backupbuff and decryption) is the same for both approaches. 
Since Access Tracker outperforms Fault Tracker, we use it for the rest of the experiments.

\begin{figure}
    \centering
    \includegraphics[width=\linewidth]{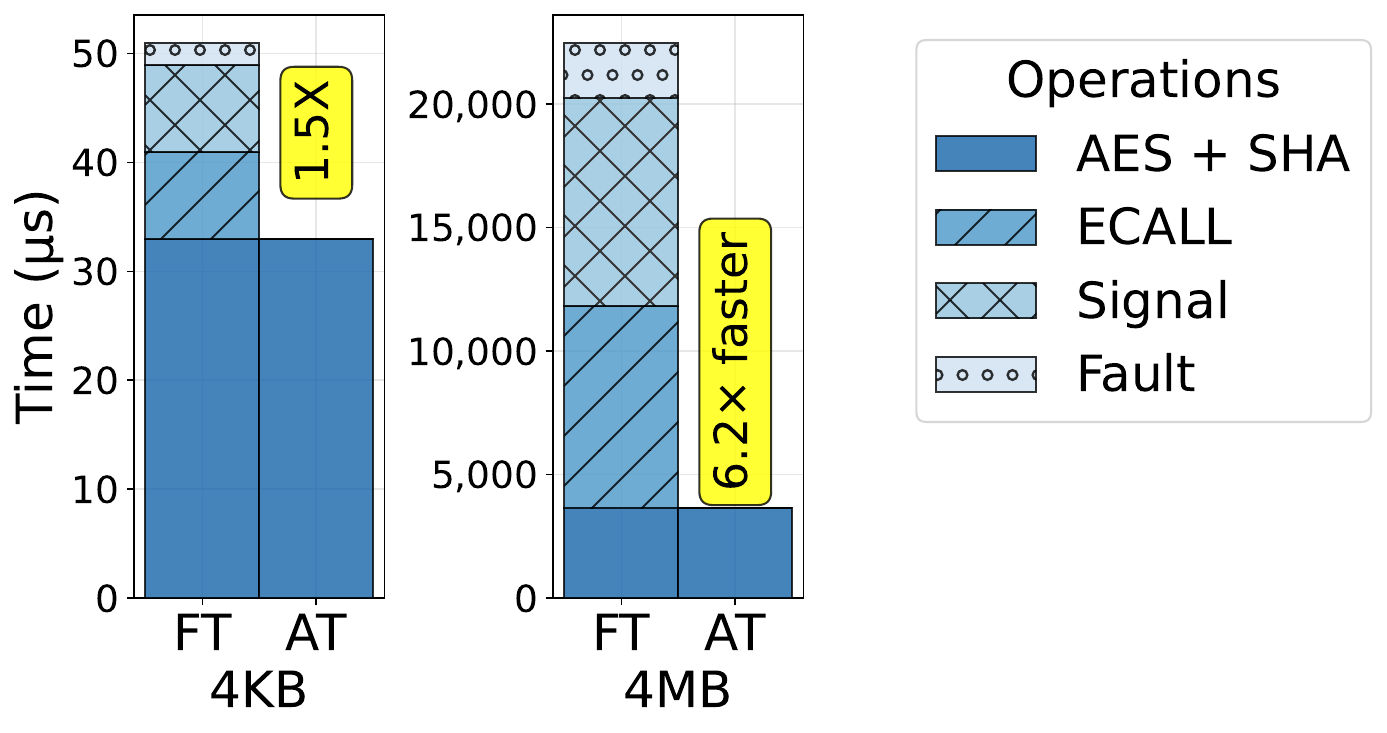}
    \caption{Performance of access tracker and fault tracker approach for consistency checker for 4\,KB and 4\,MB data. As fault tracker needs to fault on each page to restore it, the overhead of ensuring consistency checker is much higher compared to access tracker. FT: Fault Tracker, AT: Access Tracker}
    \label{fig:fault_tracker_vs_access_tracker}
\end{figure}

\sepblock
{\bf Post Migration}
After migration, Access Tracker has a small runtime overhead, which Fault Tracker does not have (\S\ref{sec:eval_additional_checks}).

\section{Security Analysis}
\label{sec:security_analysis}

Table~\ref{tab:trust_model} lists the key components of \methodname along with their trust level and the data or access pattern accessible to them.
The design and implementation of \methodname ensures protection against the following attacks.

\sepblock
\textbf{Replay attack:}
An adversary might intercept messages transmitted during migration and attempt to replay them to migrate the application to a compromised destination. 
Any attempt to replay stale data will result in an integrity check failure during restoration (\S\ref{sec:design_async_restore}) and will lead to
an immediate abort of the migration process (see \textit{Invariant 4}). Replaying the data during the same migration session will result in the restore thread dropping the packet after decryption, as the restoration of the replayed page is already done (see \textit{Invariant 3}).


\begin{table}[t]
    \footnotesize
    \centering
    \begin{tabular}{R{2.6cm}|c|L{2.8cm}}
         Entity& Trusted? & Access  \\ \hline
         Operating system &\xmark & Page fault sequence \\ \hline
         Migration manager (CRIU) & \xmark & Page fault sequence \\ \hline
         Trusted Library (\trustedLib) & \cmark &  Everything \\ \hline
         Save thread (\savethread) & \cmark & Everything \\ \hline
         Restore thread (\restorethread) & \cmark & Everything \\ \hline
         
    \end{tabular}
    \caption{Entities involved in migration and their trust level.}
    \label{tab:trust_model}
\end{table}

\sepblock
\textbf{Forking attack:}
In a forking attack, the application is maliciously
\textit{forked} during a migration. Since \masterkey is delivered only once, the forked instance will be unable to decrypt the pages, effectively preventing the attack.
Moreover, \methodname does not allow an application to resume on the source machine after the key transfer, preventing re-migration to any other destination (see \textit{Invariant 2}).

\sepblock
\textbf{Changing the Fault Address:}
An attempt by a malicious OS to disrupt the execution flow by changing the faulting virtual address from $VA_A$ to $VA_B$ will not result in any security leak.
Based on the manipulated faulting request, the contents of the page $VA_B$ will be retrieved from the source machine in an encrypted form, 
which is guaranteed to be genuine (Invariant 1). 
Since the encrypted packet contains the address $VA_B$ (\S\ref{fig:design_ecall_save}), the destination will write the page contents to address $V_B$, which is \underline{correct}. 
Of course, the application will again immediately fault as the page fault for the address $VA_A$ is yet to be serviced. However, the point to note is that there is no security violation.

\newcommand{\ack}{\texttt{ACK}\xspace}
\subsection{Fault Tolerance}
\label{sec:fault_tolerance}
\sepblock
  \textbf{Fault Tolerance}
A migration process may fail for several reasons: source crashes, destination crashes, network failures, etc.
What happens in such a situation depends on the security guarantees of the environment.
To handle such cases, we extend the proposal by ~\cite{mig_live_small_mm_prob2}.
The original enclave (OE) and restored enclave (RE) can communicate securely via remote attestation. 
\one RE, on receiving \masterkey, sends a message to OE. 
The \savethread requires the message to start saving encrypted data in \backupbuff. 
If OE does not receive the message, it times out and resumes. 
Even if RE is executing, it will be a fresh instance of the enclave as it will \textit{not} receive any data from OE. \two If it does receive the message, it is aware of two things: RE has received \masterkey and has resumed. OE \textit{cannot} be resumed now. If, for any reason, RE crashes, the application instance is lost. 
\three Once \savethread is done, it zeros all the secure pages and kills OE (since all the data has been saved). 

\section{Evaluation}
\label{sec:evaluation}
We evaluate two configurations of \methodname: \textit{\optmigvo}, where the enclave is initialized with the maximum required memory upfront, and \textit{\optmigvt}, which leverages SGX's EDMM capabilities to initialize with a smaller committed memory footprint and dynamically add pages as needed~\cite{edmm_managemetn_systor,tug_of_war}.
We compare \methodname with the current state-of-the-art MigSGX~\cite{migsgx_mm}. All other prior work~\cite{migsgx_ctr,mig_containers_mm,mig_teender_hsm_mm,mig_replicatee_consensus_mm} is based on a similar migration scheme. We also evaluate \tdxstar -- TDX with no migration due to the non-availability of two machines with TDX capability. We only report the initialization time for a CVM on the destination machine -- a \textit{lower bound} on the total downtime, as there will be other costs, such as save, transfer, and page faults~\cite{migtd_gitrepo}.

\noindent
	extit{Workloads:} Table~\ref{tab:workloads} lists the workloads used for evaluation. 
Most of the workloads in our evaluation setup either do not issue a lot of \texttt{free()} calls and hence are unaffected by the ratio of committed to uncommitted memory~\cite{tug_of_war, edmm_managemetn_systor}.
Hence, we initialize most of the workloads with $\approx$ 100\,MB to 200\,MB of committed memory with the exception of \openssl. \openssl, within an enclave, allocates memory, performs encryption of a file, and frees the memory. It repeats the process for every new request. Hence, it is initialized with $\approx 460$\,MB of committed memory.


\begin{table}[t]
    \centering
    \footnotesize
    \caption{Description of the workloads.}
    {%
        \begin{tabular}{|L{1.7cm}|p{4.1cm}|C{1.7cm}|}
            \hline
            \textbf{Workloads}                        & \textbf{Description}                                                               & \textbf{Input/Setting}                    \\ \hline
            \bfs~\cite{ligra}                         & \scriptsize{Traverse graphs generated by web crawlers. Use breadth-first search.}  & Nodes 150\,K Edges 1.9\,M \\ \hline
            \btree~\cite{btree}                       & \scriptsize{Create a B-Tree and perform lookup operations on it.}                  & Elements: 16\,M \\ \hline
            \hashjoin~\cite{hashjoin}                 & \scriptsize{Probe a hash-table (used to implement equijoin in DBs)}                &Lookups: 20\,M  \\ \hline
            \kvs~\cite{faas_kvs_cloudburst} & \scriptsize{Read and write operations on a key-value store.}             & Elements: 90\,M \\ \hline
            \openssl~\cite{openssl}                   & \scriptsize{Encryption-decryption library.}                                        & File size: 1.1\,GB              \\ \hline
            \pagerank~\cite{ligra}                    & \scriptsize{Assign ranks to pages based on popularity (used by search engines).}   & Nodes 5000
Edges 12.5 M \\ \hline
            \svm~\cite{svm}                           & \scriptsize{Popular ML algorithm (application: text and hypertext categorization)} & Rows 10000 Features 128  \\ \hline
        \end{tabular}%
    }

    \label{tab:workloads}
\end{table}

\begin{figure*}[!t]
    \centering
    \begin{subfigure}[b]{.33\linewidth}
    \includegraphics[width=\linewidth]{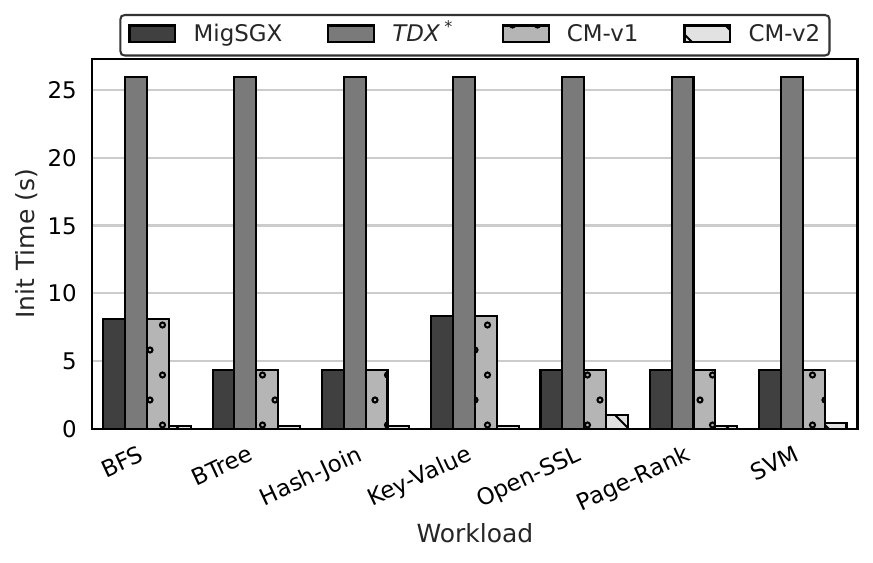}
    \caption{Total initialization time}
    \label{fig:end_to_end_init_time}
    \end{subfigure}
    \begin{subfigure}[b]{.33\linewidth}
    \includegraphics[width=\linewidth]{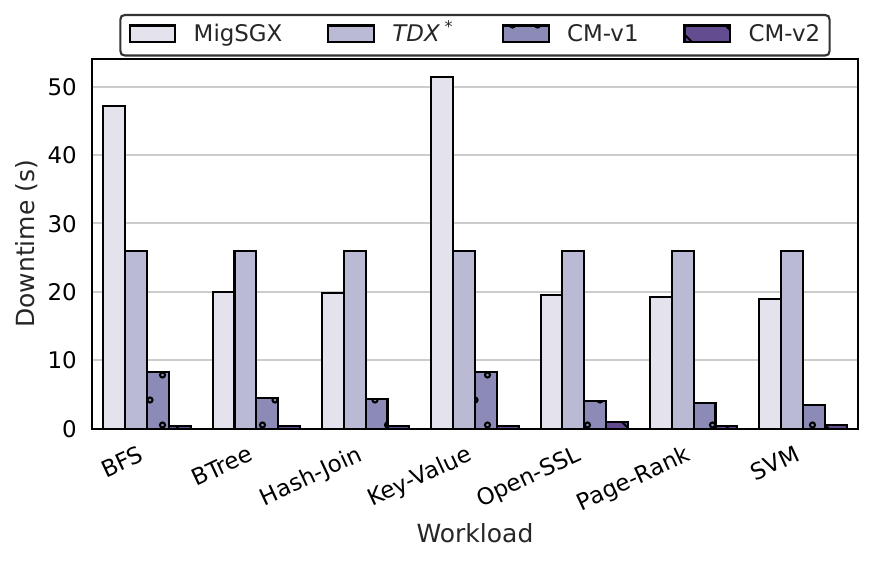}
    \caption{Total downtime}
    \label{fig:end_to_end_downtime}
    \end{subfigure}
    \begin{subfigure}[b]{.33\linewidth}
    \includegraphics[width=\linewidth]{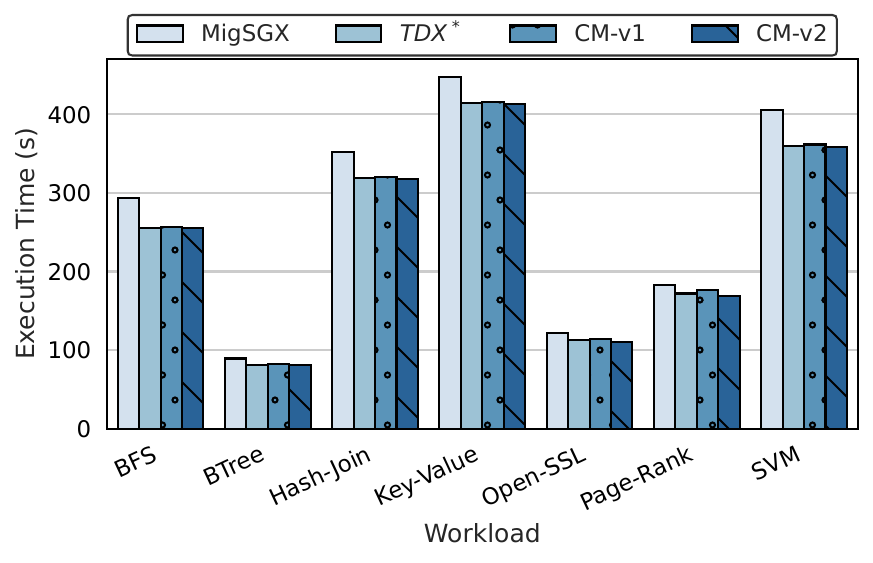}
    \caption{Total execution time -- single migration}
    \label{fig:end_to_end_exec_time}
    \end{subfigure}
     \caption{Figure showing end-to-end performance overhead due to migration with \migsgx, \tdxstar, \optmigvo (CM-v1), and  \optmigvt(CM-v2). Baseline is SGX + no migration.}
    \label{fig:end_to_end_results}
\end{figure*}

\subsection{Experimental Setup}
\label{sec:eval_setup}
We use two 2-socket systems, each with two Intel Xeon Gold 6354 CPUs (72 cores), 1TB of memory, 512GB of SSD, and 1\,TB of HDD storage. The system runs the Linux kernel version 5.15. A 1 Gbps network connects the systems. Both the systems support scalable \sgx with an EPC of up to 64 GB. We use Intel SGX SDX v2.13 and the SGX driver~\cite{intelsgxdriver} for our \sgx setup.
We use \criu (Checkpoint and Restore In Userspace)~\cite{criu} as our migration manager and use LLVM for the compiler pass. Table~\ref{tab:workloads} shows the workloads used to evaluate \methodname.  For \tdxstar evaluation, we use a machine with an Intel Xeon Gold 6526Y CPU, 256GB of main memory, and 2\,TB of HDD storage.  Our changes are mainly limited to the SGX driver and SDK.

\subsection{Evaluation -- Single Migration}
Here, we evaluate the impact of different migration schemes when an application is migrated just once.

\subsubsection{Total Downtime} 
\label{sec:eval_downtime_perf} 
Figure~\ref{fig:end_to_end_init_time} and Figure~\ref{fig:end_to_end_downtime} show the total enclave and CVM initialization time and the total downtime experienced by the workloads for a single migration for \migsgx, \tdxstar, \optmigvo, and \optmigvt. The asynchronous heap migration in \methodname and \tdxstar, which removes it from the critical path, translates to a significant reduction in the total downtime of an enclave/CVM.
\optmigvo reduces the total downtime when compared with \migsgx and \optmigvo by up
to 99.4\% and 96.2\%, respectively. The additional performance gains in \optmigvt are due to a reduction in the time of initialization of a fresh enclave on the destination machine by reducing the committed memory in \optmigvt.
The impact on the final execution time for the workloads is shown in Figure~\ref{fig:end_to_end_exec_time}.
\optmigvt outperforms \migsgx and \optmigvo by an average of $\approx10\%$ and $\approx2\%$, respectively. However, the performance difference widens for a long running application with multiple migration (see \S\ref{sec:eval_at_scale_exps}).


\subsubsection{End-to-End Performance (Single Migration)}
\label{sec:end_to_end_performance}
Efficient migration also helps in improving the end-to-end performance, as the application is resumed almost instantly on the destination.
In terms of the total end-to-end performance of the application being migrated.
\optmigvo outperforms \migsgx by up to $\approx$12.61\%. \optmigvt outperforms \migsgx by up to $\approx$12.95\%. The end-to-end performance gain is due to a quick resumption of the application on the destination machine for \optmigvo and \optmigvt. 
Note that in \migsgx, once data restoration is complete, the application runs without any overhead.
However, in \methodname, even though the application starts responding almost immediately, there are performance overheads due to a background data restore and network page faults. In spite of this, \methodname outperforms in terms of end-to-end performance.

\subsection{At-scale -- Multiple Migrations}
\label{sec:eval_at_scale_exps}
Here, we measure the performance impact of different migration mechanisms when a long-running \btree workload with a working set of 40\,GB is migrated multiple times.

\begin{figure}
    \centering
    
    \includegraphics[width=\linewidth]{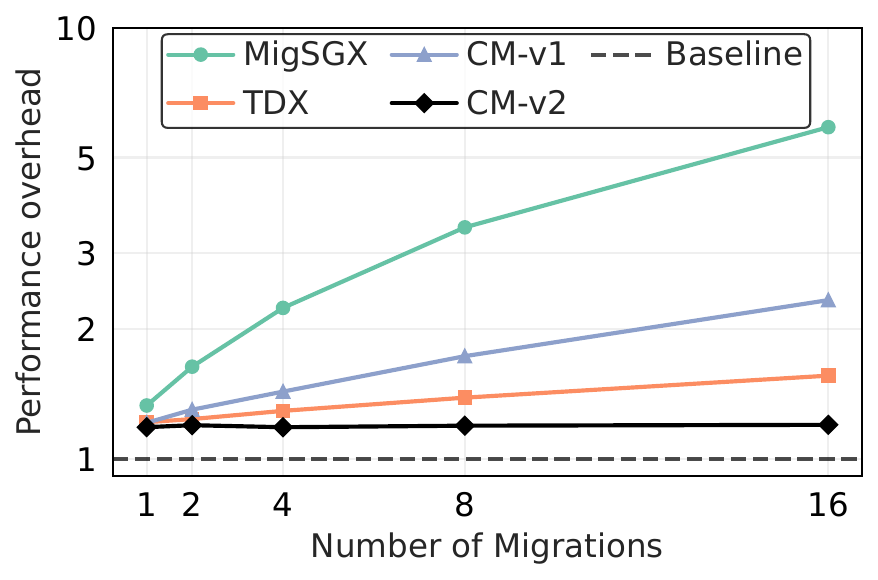}
    \caption{Performance overhead in \btree with different number of migrations.}
    \label{fig:btree_exec_time_vs_migrations}
\end{figure}

As shown in Figure~\ref{fig:btree_exec_time_vs_migrations}, the total execution performance improvement for \optmigvt scales with the number of migrations. For a single migration, \optmigvt outperforms \migsgx, \tdxstar, \optmigvo by $\approx 11\%$, $2.6\%$, and $2.3$\%, respectively. However, for a total of 16 migrations, \optmigvt outperforms \migsgx, \tdxstar, \optmigvo by $\approx 79.6\%$, $23.1\%$, and $48.6$\%, respectively. The scaling in performance improvement is due to the low downtime of the \optmigvt whereas others have a large downtime which adds up with each migration.
Note that the migrations are done one after another, there are no ``nested" migrations.
 
\subsection{No Migration Performance} 
\label{sec:eval_additional_checks} 

\begin{figure}
    \centering
    \includegraphics[width=.9\linewidth]{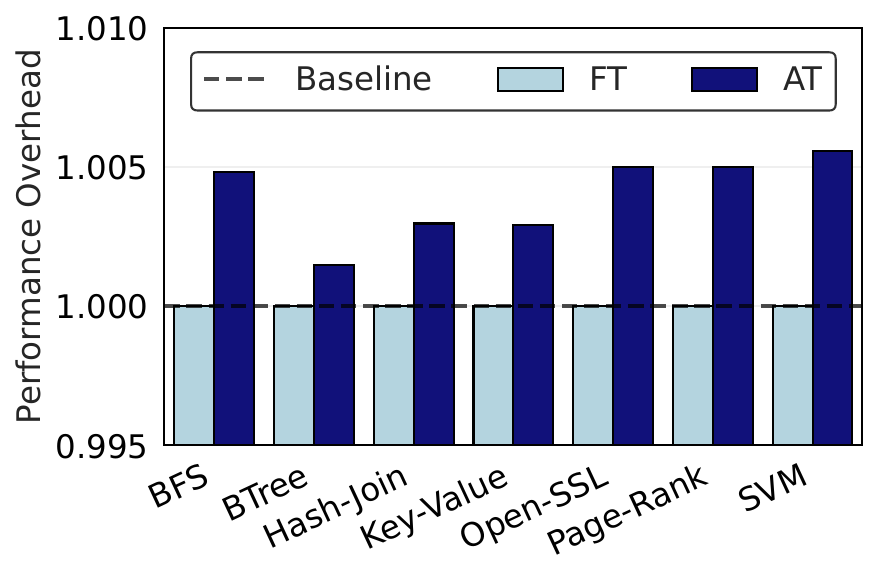}
    \caption{Figure showing performance overhead when there is no active migration. FT: Fault Tracker, AT: Access Tracker}
    \label{fig:no_migration_ovh_normalized}
\end{figure}

Migrations in data centers, if done, are not a frequent operation. Hence, we need to ensure low overhead when the application is not migrating (or is done migrating). 
Figure~\ref{fig:no_migration_ovh_normalized} shows the overhead experienced by workloads with \textit{access tracker} and \textit{fault tracker}. In the access tracker, the additional checks return immediately. There is undoubtedly a small overhead involved; however, a lot of this gets hidden because of the predictability of the branch and the unused resources of out-of-order Intel CPUs.
The performance overhead with access tracker is limited to 0.1\% to 0.6\%.
However,  with fault tracker, there is no performance overhead as there are no faults generated once the migration is done (or not yet started).

\subsection{Deep Dive -- Impact on Throughput} 
\label{sec:eval_throughput} 
We use a reference
key-value store (or KVS) workload~\cite{migsgx_mm} to report the observed throughput before, during, and
post-migration (\migsgx has done the same). We compare the throughput of \methodname with \sota.  The KVS workload performs a fixed number of {\em
set}
and {\em get} operations in a single thread with a value of size 10 KB. It reports the throughput that it observes. We report
the throughput observed before, while, and after the migration. 
As shown in Figure~\ref{fig:thp}, the throughput goes to zero when the application is frozen. 
The total
downtime is much lower in \methodname as compared to 
\migsgx (84\% and 96\% for \optmigvo and \optmigvt, respectively).
Furthermore, we observe that post-migration, the observed throughput for \optmigvt is the same as that of native execution, whereas \optmigvo suffers a nominal overhead of $0.1\%$.

\begin{figure}
    \centering
    \includegraphics[width=.9\linewidth]{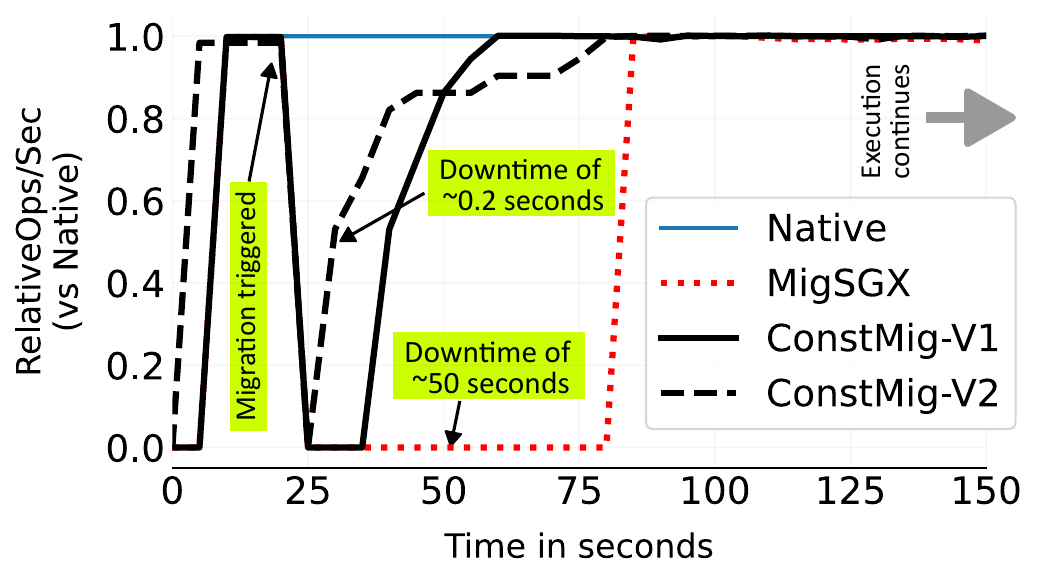}
    \caption{{Figure showing the drop in the throughput of \kvs when migrating the application with three different techniques: \migsgx, \methodname-V1, and \optmigvt.}
    }
    \label{fig:thp}
\end{figure}

\subsection{Network Page Faults (NF) }
\label{sec:eval_nf}
If the restored application tries to access the to-be-migrated data pages, it issues a network fault (NF). We measure the latency of a single network fault to be around 4-10$\mu$s.
For workloads with a sequential access pattern~\cite{securefs} or a tightly compressed organization of data (adjacent nodes of a tree are present in the same 4K page or nearby pages), there will be either few or no NFs.
However, there will be a lot of network page faults for workloads that access data randomly or where the data is not packed so tightly. This is because the application accesses will be spread out across the allocated address space. The table below shows the total network faults for three workloads: \bfs, \hashjoin, and \svm (highest network fault count).

\begin{center}
	\begin{tabular}{l|llll}
		\centering
		Workload & BFS                        & HashJoin                  & SVM                        & Rest \\ \toprule
		\#NF     & {10,064} & {1,874} & {12,308} & $<1000$    \\ \hline
	\end{tabular}\par
 \end{center}

\bfs, a graph search algorithm can reach any node in the tree with few hops. Similarly, in \hashjoin, the hash table can be randomly indexed based on the key and the hash value. In \svm, the organization of the trained model is such that it needs to access different addresses scattered in the address space. Hence,
they have a high network fault rate (consequence of low spatial locality).

\sepblock
\textbf{Worst-case scenario:} In the worst-case scenario, during migration, every page access on the destination machine causes a network fault. In this case, the total migration time for \methodname may exceed that of \migsgx. In such extreme and unlikely cases, a \stopandcopy mechanism is suited as the application would anyway be blocked on data to be restored.

\section{Related Work}
\label{sec:related_work}
Table~\ref{tab:related_work} presents an overview of the prior work in this area.  Gu et
al.~\cite{mig_live_small_mm_prob2} and Yuhala et al.~\cite{mig_plinius} use the \stopandcopy method to migrate \sgx applications from one machine to another.  The former focused on the efficient transfer of keys using a remote attestation process, while the latter focused on resuming machine learning applications in case of an application crash.  Nakashima and Kourai~\cite{migsgx_mm} and Nakatsuka et al.~\cite{migsgx_ctr}, in their work \textit{MigSGX} and \textit{CTR}, respectively, 
point out that the enclave size can be in GBs, and using the \stopandcopy method might require a large amount of memory. However, their main contribution is in
reducing the size of the unsecure buffer required to transfer the enclave data from one machine to another by employing pipelined memory transfer. Their scope is far more limited as compared to our work that looks
at full end-to-end migration.

\begin{table}[!htb]
    \caption{A summary of related work}
    \label{tab:related_work}
    \footnotesize
    \centering
    {%
    \begin{tabular}{|l|L{1.1cm}|L{2cm}|l|l|}
        \hline
        \textbf{Name}                                 & \textbf{SDK} & \textbf{Method}           & \textbf{Memory size} \\ \hline
    CloudMig ~\cite{mig_live_small_mm_prob2}      & Custom       & Stop-and-Copy             & $\approx$32MB        \\ \hline
    Plinius~\cite{mig_plinius}                    & Simulation   & Stop-and-Copy \& NVMM     & MBs                  \\ \hline
    MigSGX~\cite{migsgx_mm}                       & Intel        & Stop-and-Copy \& Pipeline & GBs                   \\ \hline
    CTR~\cite{migsgx_ctr}                       & Open Enclave        & Stop-and-Copy \& Pipeline & GBs                  \\ \hline \hline
    \rowcolor{LightCyan} \textit{\methodname}     & Intel        &        Post-copy                &            GBs          \\ \hline
        \end{tabular}
    }
    \end{table}

\section{Conclusion}
\label{sec:conclusion} 

In this paper, we present \methodname -- an efficient migration mechanism for large \sgx enclaves. We first highlight the scalability limitations of existing approaches, which still rely on the decade-old \stopandcopy migration technique. To address these challenges, we design a secure and efficient save-and-restore mechanism that enables asynchronous migration of enclave data, ensuring constant downtime during migration, regardless of enclave size.

Furthermore, we introduce two novel techniques to prevent the use-before-restore issue in restored enclaves: (i) overloading the \sgx page-fault handler and (ii) augmenting each memory access with an additional integrity check. We implement \methodname and evaluate it using real-world applications. Our results show that \methodname reduces total downtime by up to 96\% for large enclaves compared to the state-of-the-art.

\balance
\bibliographystyle{ACM-Reference-Format}
\bibliography{refs,new_refs}

\newpage
\clearpage
\balance
\nobalance
\appendix
\section{Artifact for Dirty Bit Evaluation}
\label{appendix:dirty_bit}

The source code for evaluation can be downloaded from here: \url{https://zenodo.org/records/14677356}.

\subsection{Prerequisite}
The system should satisfy the following hardware and software requirements:
\begin{enumerate}
\item An Intel SGX machine with Intel SGX SDK and Intel SGX platform software installed.
\item Python
\item Root privileges to access the pagemap.
\end{enumerate}

\subsection{Code Structure}
The evaluation codebase consists of the following files:
\begin{enumerate}
\item py\_tracker.cpp: Tracks the dirty bits of a given pid and address ranges using the pagemap interface.
\item normal\_app.cpp: Traditional non-SGX application.
\item SecureApp/: A directory that contains an SGX-based application.
\end{enumerate}

To run the py\_tracker.cpp file in the background, the user must have root permissions. When executing the normal\_app.cpp file, the user should observe that the tracker successfully tracks the dirty memory pages. However, when the secure application is executed, the tracker will be unable to detect dirty memory pages.

\subsection{How to run }
Please refer to the README file in the code.

\subsection{Monitoring PageMap}
The code in Listing~\ref{python_dirty} shows the code that opens the pagemap file from the 
{\em proc} interface for a given {\em pid} and reads the dirty bits from the specified address range.

\begin{lstlisting}[language=python,label=python_dirty,caption=Reading of dirty bit given a pid and address range., numbers=none]
...
pagemap_path = f'/proc/{pid}/pagemap'
    with open(pagemap_path, 'rb') as f:
        f.seek(start_page * 8) # Each entry is 8 bytes
        for page in range(start_page, end_page):
            entry = f.read(8)
            if not entry:
                break
            pte = struct.unpack('Q', entry)[0]
            soft_dirty = (pte >> 55) & 1  # Check bit 55
...
\end{lstlisting}

\end{document}